\documentclass{elsart}
\usepackage{amssymb}
\usepackage{amsmath}
\usepackage{graphics}

\usepackage{epsfig}
\usepackage{amssymb}

\begin{document}

\begin{frontmatter}

\title{ 
{\bf $^3$He Structure and Mechanisms of   $p^3$He  Backward
 Elastic Scattering }
  }

\author[dubna]{Yu.N.~Uzikov}
\footnote{E-mail address: uzikov@nusun.jinr.ru},
\author[ikp]{J.~Haidenbauer}
\vskip 5mm
\address[dubna]{Joint Institute for Nuclear Research, LNP, 141980 Dubna,
 Moscow Region, Russia}
\address[ikp]{Institut f\"ur Kernphysik, Forschungszentrum J\"ulich, 52425
 J\"ulich, Germany}


\begin{abstract}
{The mechanism of $p^3$He backward elastic scattering is studied. 
 It is found that the triangle diagrams with
 the subprocesses $pd\to ^3$He$\,\pi^0$,
 $pd^*\to ^3$He$\,\pi^0$ and $p(pp)\to^3$He$\,\pi^+$, where $d^*$ 
 and $pp$ denote the singlet deuteron   and diproton pair 
 in the $^1S_0$ state, respectively,
 dominate in the cross section  at 0.3-0.8 GeV, and their contribution
 is comparable with that for
 a sequential transfer of a $np$ pair at 1-1.5 GeV.
 The contribution of the $d^*+pp$, estimated on the basis of the spectator
 mechanism of the $p(NN)\to ^3$He$\,\pi$ reaction,
 increases the $p^3$He$\to ^3$He$\,p$ cross section 
 by one order of magnitude as compared to the contribution
 of the deuteron alone.
 Effects of the initial and final states interaction 
 are taken into account.

 }
\vspace{1cm}
\end{abstract}


\begin{keyword}
Elastic scattering;  Short-range interaction of nucleons

\begin{PACS}
13.75.Cs, 25.45.D, 25.10.+s\\[1ex]
\end{PACS}
\end{keyword}
\end{frontmatter}


\newpage
\baselineskip 4ex

\section{Introduction}
 
 Over the past few years $p^3$He backward elastic scattering  
 has been investiged \cite{ladouz92,blu96,uznpa98}
 on the basis of the DWBA method
 using a trinucleon bound-state wave function \cite{bkt} 
 obtained from solving the Faddeev
 equations for the RSC nucleon-nucleon ($NN$) potential. 
 Those studies suggest that this process at beam
 energies $T_p=1-2$~GeV can give unique information
 about the high momentum component of the $^3$He wave function
 $\varphi^{23} ({\bf q}_{23},{\bf p}_1)$, and specifically for
 high relative momenta, $q_{23}>0.6$ GeV/c, 
 of the nucleon pair $\{23\}$ in the $^1S_0$ state
 and low momenta of the nucleon "spectator"  $p_1<0.1$ GeV/c.
 Here $\varphi^{23}$ is the first Faddeev
 component of the full wave function of $^3$He,
 $\Psi(1,2,3)=\varphi^{23}+\varphi^{31}+\varphi^{12}$.
  The calculations presented in Refs. 
\cite{ladouz92,blu96,uznpa98}
 demonstrate a dominance  of 
 the mechanism of sequential transfer (ST) of the
 proton-neutron pair (Fig.\ref{ope}a) 
in this process over a wide range of initial energies
 $T_p=0.1-2$ GeV, except for the region of the ST dip at around 0.3 GeV.  
 Other mechanisms of two-nucleon transfer,
 such as  the deuteron exchange \cite{abdelmonem},
 non-sequential $np$ transfer \cite{blu96},
 and direct $pN$ scattering \cite{landau,uzechaya98}
 involve very high internal momenta in the $^3{\rm He}$ wave function
in $q_{23}$ as well as in $p_1$ and, in sum, give much smaller 
 contributions. However, in analogy
 to $pd$ backward elastic scattering \cite{graigwilkin69}
 one should expect also a significant contribution from mechanisms 
 related to excitation of nucleon isobars in the intermediate
 state followed by emission of virtual pions.   
  Such mechanisms were discussed in Refs. 
 \cite{uznpa98,barry73}
  on the basis of the triangle diagram of the one pion exchange (OPE)
  with the subprocess $pd\to\,^3{\rm He}\,\pi^0$ (Fig.~\ref{ope}b) and 
   in Ref. \cite{berthet81,kobap} for the two-loop diagram 
   with the subprocess $\pi d\to \pi d$.
  The  energy dependence of the cross section 
for $p^3{\rm He}\to\,^3{\rm He}p$ and also its absolute value 
 were explained to some extent in Refs. 
\cite{berthet81,kobap}. However, a common drawback of the models 
 \cite{barry73,berthet81,kobap} is the neglect of
 both (i) the contribution of the 
 singlet deuteron $d^*$ 
(i.e. where the $pn$ pair is in the spin-singlet ($^1S_0$) state) 
 in $^3$He and (ii) distortions 
 coming from  rescattering in the initial and final states. 
 In the present paper we consider both these effects 
 and show that each of them is very important, though there is an
 effective cancellation between them in the unpolarized cross section.

\section{The One Pion Exchange model}

 To account for the OPE mechanism (Fig.~\ref{ope}b-d), 
 we proceed here from the formalism of
 Ref. \cite{uznpa98} which takes into account 
 the two-body $d+p$ configuration of $^3$He.
 The $d+p$ configuration of $^3$He gives a reasonable approximation
 to the $^3{\rm He}$ charge form factor $F(Q)$ \cite{uzechaya98} 
 up to rather high transferred momenta $ Q \approx 1.5$ GeV/c.
 Furthermore, by neglecting off-shell effects in the subprocess
 $pd\to\,^3{\rm He}\,\pi^0$  one can express
 the cross section of $p^3$He scattering through the experimental
 cross section of the reaction $pd\to\,^3{\rm He}\,\pi^0$ without
 elaboration of its concrete mechanism. 

\subsection{ The mechanism of the $pd^*\to ^3$He$\,\pi^0$ reaction}

 In the calculation of Ref. \cite{uznpa98} 
the $d^*+p$ configuration was not taken into 
 account explicitly but via normalization of the form factor to $F(0)=1$. 
 In order to calculate the contribution of the meson production on the
 virtual singlet deuteron $d^*$ and on the diproton, i.e. on the
 $pp$ $^1S_0$ state in $^3$He, one has to use
 the $d^*+p$ and $(pp)+n$ configurations of $^3$He explicitly 
 and also one needs the cross sections of the reactions 
 $pd^*\to ^3$He$\,\pi^0$ and $p(pp)\to ^3$He$\,\pi^0$.

 Concerning the latter two, there are no direct measurements of this 
 reactions though there are experimental data on the inverse
 reactions of pion 
 capture, i.e. $\pi^{+}\,^{3}{\rm He}\to ppp$ from Ref.~\cite{hahn} and
 $\pi^{-}\,^{3}{\rm He}\to pnn$ \cite{gotta}, both being 
 kinematically complete experiments
 which cover the final state $NN$ interaction regions. 
 Unfortunately,  these data are restricted to pion energies 
 close to the threshold and contain only total cross sections.
Moreover, the $\pi^+$ and $\pi^-$ data
\cite{hahn,gotta} are not sufficient to deduce
 the  matrix element of the reaction $pd^*\to\,^3{\rm He}\,\pi^0$. 
 Nevertheless, these data \cite{gotta} show 
 that the formation of the $pn$ and $nn$ pairs in the final state 
 interaction region is dominated 
 by the spectator mechanism of pion absorption on the isosinglet 
 $NN$ pair in $^3{\rm He}$ (Fig.~\ref{ruderman}).   
 This mechanism is used here
 to calculate the cross section of the
  subprocesses $p(NN)_{s,t}\to\,^3{\rm He}\,\pi$ on
 the spin-triplet ({\it t}) and singlet ({\it s}) $NN$ pairs.
 According to Ref. \cite{lagetlec}, this two-body mechanism
 explains reasonably well the cross section
 of the reaction $pd\to\,^3{\rm He}\,\pi^0$ in the 
 forward $(\theta_\pi=0^\circ)$
 and backward $(\theta_\pi=180^\circ)$ directions 
 in the energy range $T_p\sim 0.3-1.0 $ GeV.
 Within a similar model, 
 the tensor analyzing power $T_{20}$ (at $\theta_\pi=180^\circ$)
 could be reasonably explained in Ref. \cite{germwilkin},
 but using the pure
 deuteron and singlet deuteron in the intermediate state of the diagram
 in Fig.~\ref{ruderman} instead of the $NN$ loop. 
 At higher energies, $T_p>1$ GeV,  
the spectator mechanism fails to reproduce
 the second peak in the excitation function of the reaction
 $pd\to\,^3{\rm He}\,\pi^0$.
 In this region the 3-body mechanism \cite{lagetlec} 
 is expected to be more important, since all three nucleons are active
in the $^3$He at high transfer of momentum. Nevertheless  
 the latter mechanism also underestimates 
 considerably the experimental cross section at
 $\theta_\pi=180^\circ$ \cite{lagetlec}.  

 Since the mechanism of the reaction $pd^*\to\,^3{\rm He}\,\pi^0$ is not
 established at higher energies,
 a completely microscopic description of the reaction
 $p^3$He$\to\,^3{\rm He}p$ within the OPE
 model cannot be achieved at present at $T_p>1$ GeV.
 In the present analysis of the contribution of the singlet $(NN)_s$ pairs
 we concentrate mainly on the energy region $T_p=0.3-1$ GeV,
 where the spectator diagram in Fig.~\ref{ruderman} dominates.

\subsection{ Formalism}

\begin{figure}[hbt]
\mbox{\epsfig{figure=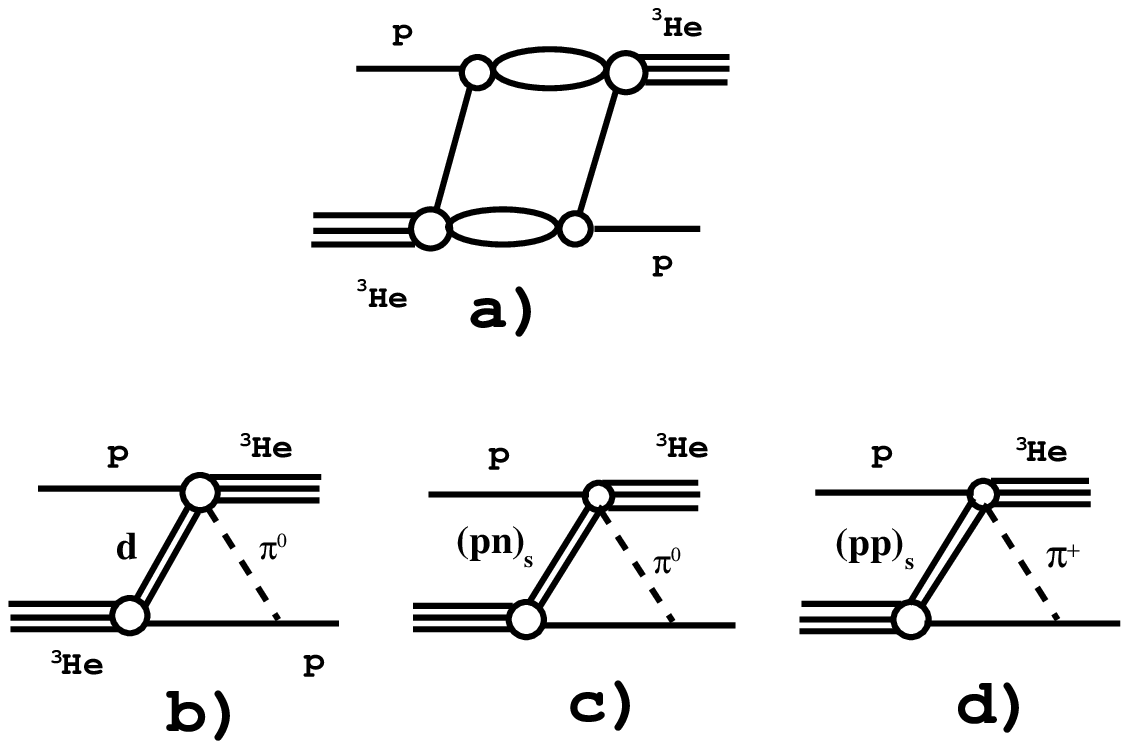,height=0.4\textheight, clip=}}
\caption{The sequential transfer (ST) ({\it a}) and one pion
exchange (OPE) ({\it b-d}) mechanisms of $p^3{\rm He}$ backward elastic
scattering with intermediate deuteron ({\it b}), singlet $pn$ pair
($d^*$) ({\it c}), and singlet $pp$ pair (diproton) ({\it d}).
}
\label{ope}
\end{figure}

 The reaction amplitude is given by 
 the coherent sum of the OPE amplitudes $M_d+M_{d^*}+M_{pp}$,
 with contributions from the deuteron $M_d$ (Fig.~\ref{ope}b), 
 singlet deuteron $M_{d^*}$ (Fig.~\ref{ope}c)
 and diproton $M_{pp}$ (Fig.~\ref{ope}d).
 For the evaluation of 
 the individual amplitudes, we use the overlap integrals 
 $^3{\rm He}-d$ and $^3{\rm He}-d^*$
 from Ref. \cite{sciavilla}.
 The $^3{\rm He}-d$ overlap wave function contains the S-wave and D-wave
 components. 
 As was shown in Ref. \cite{uznpa98}, the D-wave component of the 
 $^3$He$-d$ overlap integral is negligible in the OPE amplitude.
 Keeping the S-wave in  the $^3{\rm He}-d$ overlap wave function,
 one can find for the OPE amplitude of the reaction 
 $p^3He\to\,^3{\rm He}p$  with the subprocess $pd\to ^3He\,\pi^0$
 the following form \cite{uznpa98}
\begin{equation}
\label{oped}
 M_{d}^{\mu_h',\mu_p';\,\mu_h,\mu_p}=
-\sqrt{3} \,K G_d\, 
(10\frac{1}{2}\mu_p^\prime|\frac{1}{2}\mu_p^\prime)
\sum_\lambda (1\lambda\frac{1}{2}\mu_p^\prime|\frac{1}{2}\mu_h)
T^{\mu_h';\,\mu_p,\lambda}_d.
\end{equation}
Here
$\mu_j$ ($\mu_j')$ is the spin
projection of the initial (final) particle $j=p,h$ ($p$ denotes the proton and 
 $h$ -- the $^3$He) and $\lambda$ is the
 spin  projection of the deuteron. $T^{\mu_h';\,\mu_p,\lambda}_d$ is the
amplitude of the reaction $pd\to^3He\,\pi^0$.
 The Clebsch-Gordan coefficients are written
 in Eq. (\ref{oped}) in standard notations. 
 The dynamical and structure factors
 $K$ and  $G_d$ will be defined below.
 For the singlet deuteron there
 is only the S-wave component in the overlap integral $^3$He$-d^*$.
 Therefore, the OPE amplitude for the $d^*$ can be written as
\begin{equation}
\label{opeds}
 M_{d^*}^{\mu_h',\mu_p';\,\mu_h,\mu_p}=
 K G_{d^*}\,(10\frac{1}{2}\mu_p^\prime|\frac{1}{2}\mu_p^\prime)
 \delta_{\mu_p^\prime\, \mu_h}T^{\mu_h';\,\mu_p}_{d^*},
\end{equation}
and similarly for the intermediate diproton $(pp)$
\begin{equation}
\label{opepp}
 M_{pp}^{\mu_h',\mu_p';\,\mu_h,\mu_p}=
2\,K G_{d^*}(10\frac{1}{2}\mu_p^\prime|\frac{1}{2}\mu_p^\prime) 
 \delta_{\mu_p^\prime \, \mu_h}T^{\mu_h';\,\mu_p}_{pp}
 \end{equation}  
 where 
  $T^{\mu_h';\,\mu_p}_{d^*}$ and $T^{\mu_h';\,\mu_p}_{pp}$,
 are the amplitudes of the subprocesses 
 $pd^*\to\,^3{\rm He}\,\pi^0$ and 
 $p(pp)_s\to\,^3{\rm He}\,\pi^+$, respectively. As compared
 to Eq. (\ref{opeds}),
 an additional isospin factor of $2$ arises in Eq. (\ref{opepp})
and there is also an isospin factor of $\sqrt{3}$ in Eq. (\ref{oped}). 
 Both these factors are related only to the isospin structure
 of the lower vertices
 $^3He\to (NN)_{s,t}+N$ and $\pi\, N\to N$ of 
 the triangular diagram 
 for the OPE amplitude of the process $p^3He\to\,^3{\rm He}p$
 and do not depend
 on the mechanism of the process $p(NN)_{s,t}\to\,^3{\rm He}\,\pi$.
 The dynamical  factor $K$ is defined as
\begin{equation}
\label{kfact}
 K=
 \frac{\sqrt{m\,M (E_{p'}+m)}}{\sqrt{2\pi}\,E_{p'}}\,\frac {f_{\pi NN}}{m_\pi}
 \,  D(T_p) F_{\pi NN}(k^2). 
\end{equation}
Here  $m$, $M$ and $m_\pi$ are the masses of the proton, $^3$He 
and the pion, respectively. $E_{p'}=\sqrt{m^2+{\bf p}^2_{p'}}$ 
 and ${\bf p}_{p'}$ are the total energy and momentum of
the secondary proton in the laboratory system, and 
$f_{\pi NN}$ and $F_{\pi NN}(k^2)$ 
are the coupling constant and 
the (monopole) form factor at the $\pi NN$ vertex.
The distortion factor $D(T_p)$ is given in Ref. \cite{uznpa98} in 
eikonal approximation in terms of an analytical parametrization 
of the forward $pN$ and $p^3{\rm He}$ scattering amplitudes.

 The nuclear form factors for the    triplet  ($G_d$) and singlet ($G_{d^*}$) 
 channels are given by 
\begin{equation}
\label{gfact}
G_{d,d^*}=
\sqrt{ S_{d,d^*}^h}\left (i\kappa F_0^{d,d^*}({\tilde p})+ W_{10}^{d,d^*}({\tilde p},
{\tilde \delta})\right ).
\end{equation}
%
where
\begin{eqnarray}
\label{formfact}
{ F}^{d,d^*}_0({\widetilde p})=\int _0^\infty U^{d,d^*}_0(r)\,
 j_0({\widetilde p}r)\,
r dr,\nonumber \\
{ W}^{d,d^*}_{10}({\widetilde p},{\widetilde \delta})=
\int_0^\infty j_1({\widetilde p}r) U^{d,d^*}_0(r)(i{\widetilde {\delta }}+1)
\exp{(-i{\widetilde {\delta }}r)}dr.
\end{eqnarray}
 Here $U^{d}_0(r)$ ($U^{d^*}_0(r)$) is the S-wave component of the
 $^3{\rm He}-d$  
($^3{\rm He}-d^*$) overlap integral and $j_l$ is the spherical Bessel
 function.
The kinematical variables $\kappa$, ${\widetilde {\bf p}}$ and 
 ${\widetilde \delta}$ are  determined by the proton  momentum ${\bf p}'$
 \cite{uznpa98}:
${\bf {\widetilde p}}=2m/E_{p'}{\bf p}'$,  
$\kappa={\widetilde p}(2E_{p'}+m)/(2E_{p'}+m)$,
${\widetilde \delta}^2=2m/E_{p'}(m^2_\pi-k^2)+|{\bf {\widetilde  p}}|^2$,
 where $k^2$ is the square of 4-momentum of the virtual $\pi$-meson.     
The spectroscopic factors for the deuteron, $S_{pd}^h$, and the
singlet deuteron, including diproton, $S_{pd^*}^h$, are taken here to be
$S_{pd}^h$ = $S_{pd^*}^h$ = 1.5, in accordance with  
Refs.~\cite{germwilkin,sciavilla}. 

 We should note that due to the presence of the Kronecker-$\delta$ 
 in Eqs. (\ref{opeds})
 and (\ref{opepp}), the singlet amplitudes $M_{d^*}$ and $M_{pp}$ contribute
 only
 to the spin-independent part of the OPE amplitude of 
 the reaction $p^3{\rm He}\to\,^3{\rm He}p$. At the same time the spin-dependent
 part 
 is given by the spin-triplet amplitude $M_d$ alone.
 Because of this specific structure there is no interference between
 the triplet 
 and singlet
 amplitudes in the spin-averaged sum 
${\overline {|M_d+M_{d^*}+M_{pp}|^2}}$. This feature simplifies
 the theoretical analysis of the unpolarized cross section significantly.  
 Thus, we find that the total spin averaged OPE amplitude
  of the $p^3$He backward elastic scattering has the form 
 \begin{equation}
 \label{averagedm}   
{\overline {| M^{\mu_h',\mu_p';\,\mu_h,\mu_p}|^2}}=
 |K|^2 \left \{ {\overline {|G_d\,T_d^{\mu_h';\mu_p,\lambda}|^2}} 
 +\frac{1}{3}
 {\overline {|G_{d^*}(T^{\mu_h';\,\mu_p}_{d^*}+2\,T^{\mu_h';\,\mu_p}_{pp})|^2}}  
 \right \}.
\end{equation}


 Since there is no interference term between singlet and triplet 
 $NN$ pairs, it is convenient to introduce the following relation for
 the squared singlet and triplet  
 amplitudes of the processes $p(NN)_{s,t}\to\,^3{\rm He}\,\pi$,
\begin{equation}
\label{trisin}
\frac{1}{3} {\overline {|T_{d^*}^{\mu_h';\,\mu_p}+2T_{pp}^{\mu_h';\,\mu_p}|^2}}=
 C_I {\overline {|T_d^{\mu_h';\,\mu_p\,\lambda}|^2}}.
\end{equation}
%
Of course, in general, the factor $C_I$ is a complicated function 
depending on the energy and the transferred momentum. However,
for the spectator model (Fig.~\ref{ruderman}) we can obtain a reasonable
estimate of this factor on the basis of isospin relations only. 
We can then use Eq.~(\ref{trisin}) in order to express
the singlet contribution in terms of the experimental cross section
of the reaction $pd\to$$^3{\rm He}\,\pi^0$. 
 After that the c.m.s. cross section  
 of $p^3$He backward elastic scattering can be written as
 $$\frac{d \sigma}{d\Omega_{cm}}=
 \frac{m\,M (E_{p'}+m_p)}{2\pi\,E_{p'}^2}
\left(\frac {f_{\pi NN}}{m_\pi}
 \right )^2  \frac{s_{pd}\, q_{pd}}{s_{ph}\,q_{\pi h}}
 F^2_{\pi NN}(k^2) \left |D(T_p) \right |^2\times $$
 \begin{equation}
 \label{sec}
\times \Bigl \{ |G_d|^2 +  C_I\,|G_{d^*}|^2 \Bigr \} 
  \frac{d \sigma}{d\Omega_{cm}}(pd\to\,^3{\rm He}\,\pi^0) \ ,
\end{equation}
 where $s_{ij}$ is the square of the invariant mass of the system $i+j$,
 and $q_{ij}$ is the relative momentum in this system. 

\subsection{Approximated evaluation of $C_I$}

\begin{figure}[hbt]
\mbox{\epsfig{figure=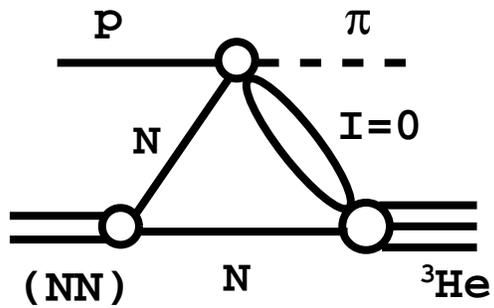,height=0.2\textheight, clip=}}
\caption{The spectator model of the reaction 
$p(NN)_{s,t}\rightleftharpoons\,^3{\rm He}\pi$. 
}
\label{ruderman}
\end{figure}

 In the evaluation of the factor $C_I$ in Eq.~(\ref{trisin}) 
 for the spectator model of the process $p(NN)_{s,t}\to ^3$He$\,\pi$
 we assume that the spatial parts of the  vertices $d\to p+n$,
 $d^*\to p+n$, and $(pp)_s\to p+p$ in Fig.~\ref{ruderman} are 
approximately the same in $^3{\rm He}$.
 Furthermore, 
 we assume that the subprocess $pN\to (NN)_t\,\pi$ dominates in 
 the upper vertex of the diagram in Fig.~\ref{ruderman} and that the  
 amplitude $pN\to (NN)_s\,\pi$ is negligible. This is true in 
 the $\Delta$-region, as was shown recently \cite{uzwilkin}.
 With this approximation
 the following relation between the amplitudes of the processes
 $pd^*\to\,^3{\rm He}\,\pi^0$ and 
 $p(pp)_s\to\,^3{\rm He}\,\pi^+$ follows from isospin invariance
\begin{equation}
\label{r2}
 T^{\mu_h';\,\mu_p}_{pp}=2\,T^{\mu_h';\,\mu_p}_{d^*}.
\end{equation}
 After that the factor $C_I$ is basically given by the
 Clebsch-Gordan coefficients at the vertices of the spectator
 diagram in Fig.~\ref{ruderman} and in the OPE diagrams in
 Fig.~\ref{ope}b-d. 
 We find that there is a constructive 
 interference between  the singlet amplitudes $M_{d^*}$ and $M_{pp}$ 
 and that the factor $C_I$ in Eqs. (\ref{trisin},\ref{sec}) equals to
 $\frac{25}{3}$. 
 Thus, the contributions of the singlet deuteron and the diproton 
 are significantly larger than those of the deuteron. 
 Since the singlet and triplet form factors are related numerically by
 $G^{d^*} \approx 1.5 \ G^d$ in the kinematical region under discussion
 (cf. Ref. \cite{germwilkin}), we get a
 total enhancement of about 12 in the $p^3{\rm He}\to\,^3{\rm He}p$ 
 cross section due to the contribution of the singlet $NN$ pairs.
 Note that we use Eq.~(\ref{trisin}) with the factor
 $C_I=25/3$ also at energies above 1~GeV. Due to the poor knowledge
 of the mechanism of the reaction $pd\to\,^3{\rm He}\,\pi$ at these energies, 
 however, this has to be considered as a purely phenomenological 
 prescription.

 In order to explore the reliability of the assumptions
 and approximations 
 discussed above, we performed also a direct calculation of the 
 term ${\overline {|T_{d^*}+2T_{pp}|^2}}$ in the left hand side of 
 Eq. (\ref{trisin}). It was done in collinear kinematics in the
 region of $T_p=0.3-0.8$~GeV on the basis of the spectator diagram 
 (Fig.~\ref{ruderman}) with the intermediate deuteron, taking into account
 the S- and D-wave components of the $^3$He-$d$ overlap integral.
 For the $pp\to d\,\pi^+$ amplitude we employed
 the parametrization 
 given in Ref. \cite{germwilk90}. As an estimation,
 for the internal wave function of the $d^*$, we used
 the separable term $\varphi_1(r)$ of the $^1S_0$ component of the
 RSC trinucleon wave function from Ref. \cite{haidukgs} and, for
 comparison, the S-wave component of the RSC deuteron wave function. 
 In both these cases 
 the obtained results for the cross section at 0.3-0.8~GeV
 coincide with the
 present estimation based on Eqs.(\ref{trisin},\ref{sec})
 within $\approx $30\%.
 At last,
 the total cross section of the reaction 
$\pi^{+}\, ^{3}{\rm He}\to (pp)p$
 in the final state interaction region measured  in Ref. \cite{hahn}
 at kinetic energy of pion 37~MeV, $\sigma=2.4\pm 0.7$~mb, is comparable with
 that for the reaction  $\pi^0\, ^3{\rm He}\to dp$, $\sigma=1.3$~mb, 
 recalculated here from the $pd\to ^3{\rm He}\,\pi^0$ data
 \cite{berthet85}  for the corresponding
 proton beam energy $T_p=262$ MeV. The ratio of these cross sections
$\sigma(\pi^{+}\,^{3}{\rm He}\to (pp)p)/\sigma(\pi^0\,^3{\rm He}\to dp)$
 is in agreement with the value $\frac{2^2}{2J_d+1}=\frac{4}{3}$,
 expected within 
 the spectator mechanism, where the factor 4 in the numerator
 is the squared isospin
 factor 2 from Eq. (\ref{r2}), and the factor 3 in the denominator 
 is the spin-statistical factor for the final deuteron. 

 The result above implies that, within this approximation, 
 the total contribution of the triplet and  singlet $NN$ pairs
 can be taken into account 
 by variation of the effective spectroscopic factor
 of the deuteron in $^3{\rm He}$, 
 $S_{pd}^h\leadsto S_{pd}^h(1+1.5C_I)$. 
 In the numerical calculation we use the parametrizations 
 from Ref. \cite{germwilkin} for the $^3{\rm He}-d$ 
 and $^3{\rm He}-d^*$ overlap integrals \cite{sciavilla} obtained for the
 Urbana $NN$ potential. We have found that the final result is almost
 the same when the RSC parametrization from Ref. \cite{uznpa98} is used.
 The experimental cross section of the reaction
 $pd\to\,^3{\rm He}\,\pi^0$ for the backward 
scattered pions ($\theta_{cm}=180^\circ$)
 is taken from \cite{berthet85}. For the cut-off parameter 
 $\Lambda_\pi$ in the monopole form factor of the $\pi NN$ vertex 
we consider values in the range of $\Lambda_\pi=0.65 -1.3$~GeV/c. 
The lower case,
 $\Lambda_\pi=0.65$~GeV/c, corresponds to the value obtained in
an analysis of the reaction $pp\to pn\pi^+$ at 0.8~GeV
 performed in the 
$\pi+\rho$ exchange model \cite{imuz88}. The upper case,
$\Lambda_\pi=1.3$~GeV/c, is the value used in the full Bonn 
$NN$ model \cite{bonn}. 

\section {Numerical results and discussion}

\begin{figure}
\mbox{\epsfig{figure=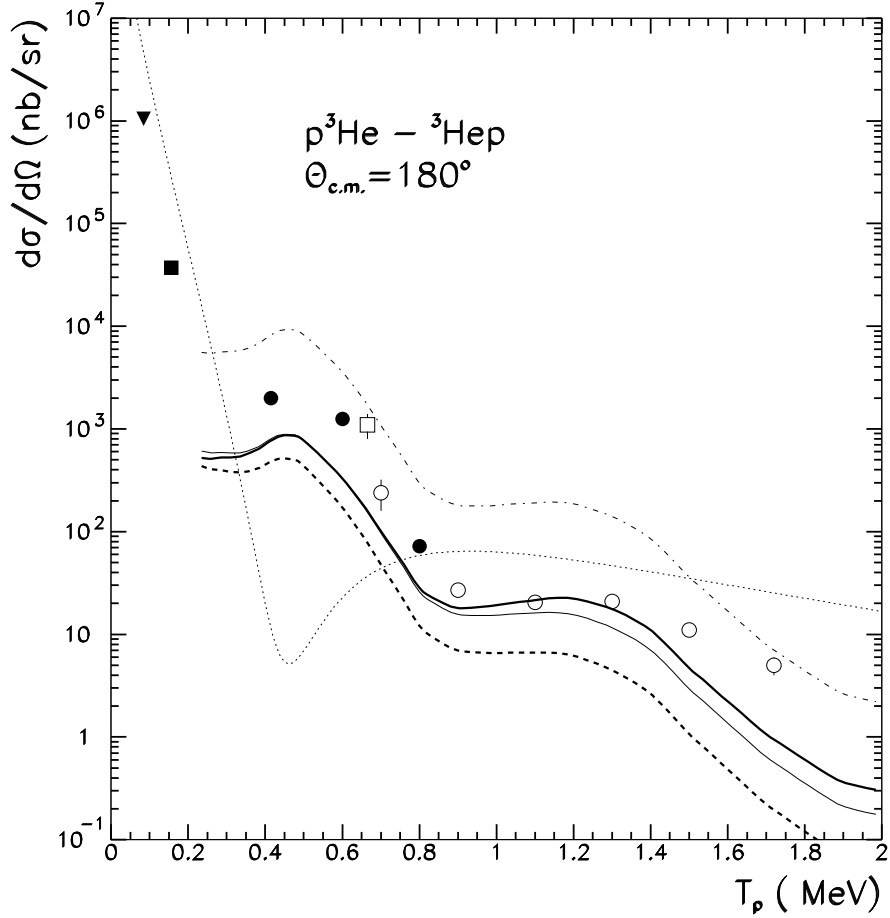,height=0.6\textheight, clip=}}
\caption{
C.m.s. cross section of elastic $p^3He$ scattering 
at the scattering angle $\theta_{cm}=180^\circ$ as a function of
the kinetic energy of the proton beam. 
 Calculations on the basis of the OPE model for the deuteron
 in the intermediate state and without distortions are shown by thin 
 solid line 
(for $\Lambda_\pi=1.3$ GeV/c) and dashed line ($\Lambda_\pi=0.65$ GeV/c). 
 OPE cross section for $d+d^*+pp$ with $\Lambda_\pi=1.3$ GeV/c
is shown by dashed-dotted (without distortions) and thick solid 
line (including distortions).
The result for the 
non-distorted ST cross section with CD Bonn is given by the dotted line. 
Experimental data are from Refs.
\protect\cite {berthet81} ($\circ$), \protect\cite{langevi} (filled square),
\protect\cite{komarov} (open square),
\protect\cite{frascaria} ($\bullet$), and \protect\cite{votta} (filled triangle).
} 
\label{cross2}
\end{figure}

 The result of our calculation are shown in Fig.~\ref{cross2}.
 One can see that the OPE model with the deuteron 
 yields a reasonable description of the energy dependence
 of the cross section for $T_p= 0.4 - 1.5$~GeV,
 although it underestimates the magnitude.
 The calculated cross section is smaller
 than the experiment by a factor of around 3--3.5 for 
 $\Lambda_\pi=0.65$~GeV/c, and by about 1.5--2.5 for
 $\Lambda_\pi=1.3$~GeV/c depending on the beam energy.
 After the contributions of the singlet deuteron $d^*$ and of the $pp$ pair
 are taken into account, the cross section for $p^3{\rm He}\to\,^3{\rm He}\,p$ 
 is overestimated at $T_p>0.3$~GeV
 by a factor of 2.5--4 (for $\Lambda_\pi$ = 0.65~GeV/c)
 and 5--10 (for $\Lambda_\pi$ = 1.3~GeV/c), respectively.
 The distortion factor $D(T_p)$ reduces the OPE
 cross section of the reaction $p^3{\rm He}\to\,^3{\rm He}\,p$
 by one order of magnitude (thick solid line) and brings it in qualitative 
 agreement with the data.  
 The discrepancy with the data in the region of the first shoulder, 
 $T_p=0.3-0.6$~GeV, can be attributed to others terms 
 in the $pd^*\to\,^3{\rm He}\,\pi^0$ amplitude, like the
 two-step mechanism \cite{konuz97}.
 It can be shown that for the two-step
 mechanism there is also an enhancement of the $d^*+p$ contribution
 in the $\Delta$-isobar region  but, in contrast to the spectator mechanism,
 its energy dependence is strongly affected by the off-shell behaviour of 
 the $\pi N$ scattering amplitude and not considered here.

\begin{figure}[hbt]
\mbox{\epsfig{figure=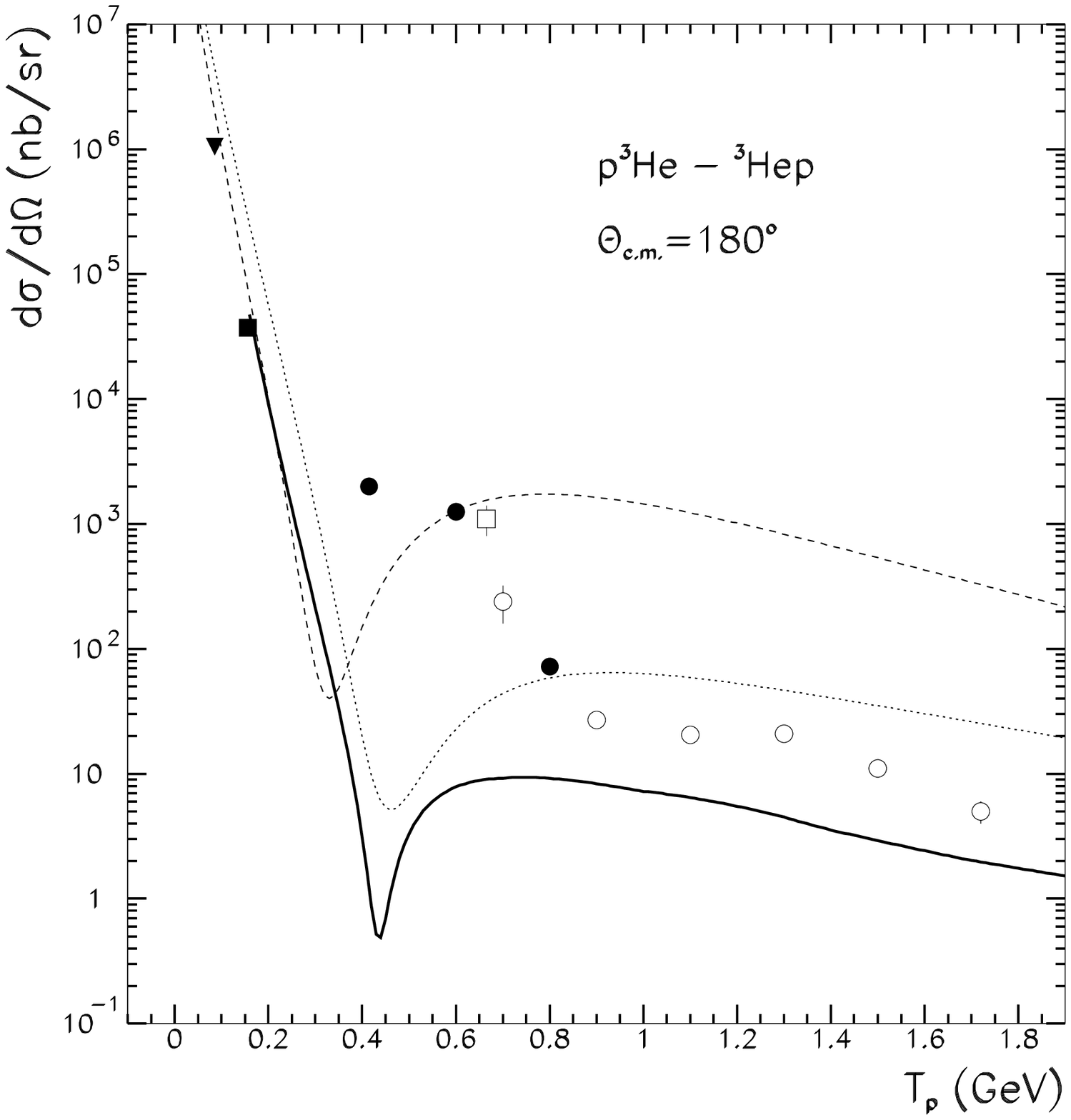,height=0.65\textheight, clip=}}
\caption{
C.m.s. cross section of elastic $p^3He$ scattering 
at the scattering angle $\theta_{cm}=180^\circ$ as a function of
the kinetic energy of the proton beam. 
The theoretical curves show results of calculations
for the ST mechanism in the Born approximation and
with different $^3$He wave functions:
Reid RSC (dashed line); CD Bonn (dotted). 
The ST cross section for the CD Bonn wave function with distortions 
taken into account is shown by thick solid line. 
Note that the distortion factor for the ST mechanism differs
from the one for OPE. 
Same description of data as in Fig.~\ref{cross2}.
}
\label{cross1}
\end{figure}

 Turning back to the pure two nucleon transfer mechanism 
\cite{ladouz92,blu96,uznpa98}
 we should note that the three-nucleon bound-state wave
 function \cite{bkt} based on the Reid RSC potential most 
 likely contains too large high momentum components
 as compared to modern $NN$ potentials.
 In order to corroborate that 
 we show here, in the framework of the S-wave formalism of
 Ref. \cite{uznpa98}, 
 that for the trinucleon wave function \cite{vbaru} based on the CD Bonn 
 $NN$ interaction \cite{cdbonn}
 the ST cross section at $T_p>0.5$ GeV is by a factor of 30 smaller
 than for the RSC. Nevertheless the predicted cross section is still
 comparable with the experimental data at 
$T_p>0.9$~GeV (Fig.\ref{cross1}).\footnote{Inclusion of the D-waves will 
 lead to an additional increase of 
 the calculated cross section (see Ref.~\cite{blu96}).}
 One can see from Fig.~\ref{cross2}, that the ST mechanism is very important
 at beam energies $T_p=0.9-1.5$~GeV and 
 it definitely dominates at low ($T_p<0.3$~GeV) and high ($T_p>1.5$~GeV)
 energies.
 A dominant role of the singlet $NN(^1S_0)$ pairs in $^3{\rm He}$,
 as reflected  in the ST and OPE
 mechanisms of the $p^3{\rm He}$ backward elastic scattering, probably,
 can be connected
 to the pp-correlations in the reaction $ ^3{\rm He}(e,e^\prime pp)n$ 
 reaction recently observed in Ref. \cite{weinstein}. 


 Summarizing our results, we can conclude the following: 
 (i) The OPE mechanism in  
 the plane wave approximation with the subprocess
 $pd\to\,^3{\rm He}\pi^0$ describes well the energy dependence of the 
 $p^3{\rm He}\to\,^3{\rm He}p$ cross section, but underestimates its absolute
 value by a factor of 2--3.5, depending on the cut-off mass used in the
 form factor at the $\pi NN$ vertex. 
 (ii) The contribution of the singlet deuteron
 for the spectator mechanism of the reaction 
 $pd^*\to\,^3{\rm He}\pi^0$ is by one order of magnitude larger than
 the one of the deuteron. (iii) The enhancement of the
 OPE cross section after inclusion of the contribution of the 
 singlet deuteron
 is, however, almost completely counterbalanced by the reduction 
 caused by distortions in the initial and final states.

 Therefore, the first shoulder in 
 the $p^3{\rm He}\to\,^3{\rm He}p$ cross section 
 at 0.4--0.6 GeV is caused mainly by the OPE mechanism with the singlet
 $NN(^1S_0)$ pairs. 
 A measurement of spin observables, planned at the RCNP in Osaka \cite{Hatan},
 can give additional information here because,
 in contrast to the $d$ term, the $d^*$ and $pp$ terms 
 contribute only to the spin-independent part of the
 OPE amplitude of the reaction $p^3{\rm He}\to\,^3{\rm He}p$ and,
 consequently, could have a strong influence on the spin-spin correlation
 parameter $C_{y,y}$.   
 The origin of the second shoulder at 0.9--1.3 GeV is less
 clear. A significant part of this cross
 section is produced by the ST mechanism \cite{ladouz92,blu96,uznpa98}.
 Using the CD Bonn wave function for $^3$He 
 instead of the one based on the RSC potential 
 decreases the contribution of the ST mechanism. 
 Nevertheless, this does not change the main conclusion of Ref. \cite{uznpa98}, 
 namely that the significance of the contribution from this mechanism
 for energies $T_p>1$~GeV allows one to probe specifically the high
 momentum components of the $^3$He wave function.
 However, the connection between the observables 
 and the high-momentum structure of the $^3$He wave function becomes much
 less transparent because of the large contribution of the OPE mechanism 
 and the uncertainties connected to its
 $d^*$ contribution in this region.
 Future progress in the analysis of the role of intermediate pions in the
 reaction $p^3{\rm He}\to\,^3{\rm He}\,p$ requires the clarification of the mechanism 
 for the subprocess $pd\to\,^3{\rm He}\,\pi^0$, in particular at energies
 $T_p>1$~GeV. In addition it is desirable to take into account 
 that this subprocess is off-shell in $p^3{\rm He}\to\,^3{\rm He}\,p$
 and also to consider the $NN$ continuum in the virtual subprocesses
 $p(NN)_{s,t}\to ^3${\rm He}$\,\pi^0$.

{\bf Acknowledgement.} We would like to thank D. Gotta and 
 C. Wilkin for fruitful comments and suggestions.
One of us (Yu.N.U.) gratefully acknowledges the warm hospitality 
at the IKP-Theory of the Forschungszentrum J\"ulich. 
This work was supported in part 
by the Heisenberg-Landau Programm of the BMBF (Germany) -- JINR (Russia) 
agreement.

\vfill 
\newpage

\eject
\eject   

\eject

\newpage
\begin{thebibliography}{99}
%
\bibitem{ladouz92} A.V.~Lado, Yu.N.~Uzikov, Phys. Lett. {\bf B279} (1992) 16.
%
\bibitem{blu96} L.D.~Blokhintsev, A.V.~Lado, Yu.N.~Uzikov, Nucl. Phys.
{\bf A597} (1996) 487.
%
\bibitem{uznpa98} Yu.N.~Uzikov, Nucl.Phys. {\bf A644} (1998) 321;
 Phys. Rev. {\bf C 58} (1998) 36.
\bibitem{bkt} R.A.~Brandenburg, Y.~Kim, A.~Tubis, Phys. Rev. {\bf C 12} (1975)
1368.
%
%
\bibitem{abdelmonem} M.S.~Abdelmonem, H.S.~Sherif, Phys. Rev. {\bf C 36}
(1987) 1900.
\bibitem{landau} M.J.~Paez, R.H.~Landau, Phys. Rev. {\bf C 29}(1984) 2267.
%
\bibitem{uzechaya98} Yu.N.~Uzikov, Elem. Chast. At. Yadr. {\bf 29} (1998) 1010
 (Part. Nucl. {\bf 29} (1998) 417).
%
\bibitem{graigwilkin69} N.S.~Craigie, C.~Wilkin, Nucl. Phys. {\bf B14} (1969) 477.
%
\bibitem{barry73} G.W.~Barry, Phys. Rev. {\bf D 7} (1973) 1441.
%
\bibitem{berthet81} P.~Berthet {\it et al.}, Phys. Lett. {\bf 106B} (1981) 465.
%
\bibitem{kobap} A.P.~Kobushkin {\it et al.}, {\tt nucl-th/0112078}. 
%
\bibitem{lagetlec} J.M.~Laget, J.F.~Lecolley, Phys. Lett {\bf  B194} (1987) 177.

%
\bibitem{germwilkin} J.-F.~Germond, C.~Wilkin, J. Phys. G {\bf 14} (1988) 181.
%
\bibitem{hahn} H.~Hahn {\it et al.}, Phys. Rev. {\bf C 53} (1996) 1074.
%
\bibitem{gotta} D.~Gotta {\it et al.} Phys. Rev. {\bf C 51} (1995) 469.
%
\bibitem{sciavilla} R.~Schiavilla, V.R.~Pandharipande, R.B.~Wiringa, Nucl. Phys.
 {\bf A449} (1986) 219.
\bibitem{uzwilkin} Yu.N.~Uzikov, C.~Wilkin, Phys. Lett. {\bf B545} (2001) 191.
%
\bibitem{germwilk90} J.-F.~Germond, C.~Wilkin, J. Phys. G {\bf 16} (1990) 381.
%
\bibitem{haidukgs} Ch.H.~Haiduk, A.M.~Green, M.E.~Sainio, Nucl. Phys. {\bf A337}
  (1980) 13. 

\bibitem{berthet85} P.~Berthet {\it et al.}, Nucl. Phys.  {\bf A443} (1985) 589;
 V.N.~Nikulin {\it et al.}, Phys. Rev. {\bf C 54} (1996) 1732. 
%
%
 \bibitem{imuz88} A.~Matsuyama, T.-S.H.~Lee, Phys. Rev. {\bf C 34} (1986) 1089;
 O.~Imambekov, Yu.N.~Uzikov, Sov. J. Nucl. Phys. {\bf 47} (1988) 695.
%
\bibitem{bonn} R.~Machleidt, K. Holinde, Ch. Elster, 
Phys. Rep. {\bf 149} (1987) 1. 
%
\bibitem{konuz97} G.~F\"aldt, C.~Wilkin, Nucl. Phys. {\bf A587} (1995) 769;
 L.A.~Kondratuyk, Yu.N.~Uzikov, JETP Lett. {\bf 63} (1996) 1.

\bibitem{vbaru} V.~Baru, J.~Haidenbauer, C.~Hanhart, J.A.~Niskanen, 
{\tt nucl-th/0207040}, Eur. Phys. J. {\bf A}, in print. 

\bibitem{cdbonn} R.~Machleidt, Phys. Rev. {\bf C 63} (2001) 024001. 
%

\bibitem{langevi} H.~Langevin-Joliot {\it et al.}, Nucl. Phys. {\bf A158} (1970) 309.

\bibitem{komarov} V.I.~Komarov {\it et al.}, Yad. Fiz. {\bf 11} (1970) 399.
\bibitem{frascaria}  R.~Frascaria {\it et al.}, Phys. Lett. {\bf 66B} (1977) 329.
\bibitem{votta} L.G.~Votta {\it et al.}, Phys. Rev. {\bf C 10} (1974) 520.

\bibitem{Hatan} K. Hatanaka, in {\it Proceedings of the Workshop MEDIUM02},
Kyushu University, Fukuoka, Japan, October 25-26, 2002, in print. 
%
\bibitem{weinstein} L.B. Weinstein, R. Niyazov, {\tt nucl-ex/0209014}.
\end{thebibliography}
\end{document}